\documentclass[aps,prc,floats,floatfix,preprint,superscriptaddress,nofootinbib]{revtex4}   
\usepackage{graphicx}   
\def\be{\begin{equation}}  
\def\ee{\end{equation}}  
\begin{document}   
  
\title{Time-dependent density functional theory for strong   
electromagnetic fields in crystalline solids}
   
\author{K. Yabana}
\affiliation{   
Center for Computational Sciences,   
University of Tsukuba, Tsukuba 305-8571, Japan }   
\affiliation{   
Graduate School of Pure and Applied Sciences, University of Tsukuba,   
Tsukuba 305-8571, Japan}   
\author{T. Sugiyama}
\affiliation{   
Graduate School of Pure and Applied Sciences, University of Tsukuba,   
Tsukuba 305-8571, Japan}   
\author{Y. Shinohara}
\affiliation{   
Graduate School of Pure and Applied Sciences, University of Tsukuba,   
Tsukuba 305-8571, Japan}   
\author{ T. Otobe}  
\affiliation{Advanced Photon Research Center, JAEA, Kizugawa, Kyoto  
619-0215, Japan}  
\author{ G.F. Bertsch}   
\affiliation{  
Department of Physics and Institute for Nuclear Theory, University of Washington,  
Seattle 98195, U.S.A  
}  
   
\begin{abstract}   
We apply the coupled dynamics of time-dependent density functional 
theory and Maxwell equations to the interaction of intense laser pulses 
with crystalline silicon.  As a function of electromagnetic field intensity, 
we see several regions in the response.  At the lowest intensities, 
the pulse is reflected and transmitted in accord with the dielectric 
response, and the characteristics of the energy deposition is consistent 
with two-photon absorption.   The absorption process begins to deviate
from that at laser intensities $\sim10^{13}$ W/cm$^2$, where the
energy deposited is of the order of 1 eV per atom.  
Changes in the reflectivity are seen as a function of intensity.
When it passes a threshold of about $3\times 10^{12}$ W/cm$^2$, 
there is a small decrease. At higher intensities, above
$2 \times 10^{13}$ W/cm$^2$, 
the reflectivity increases strongly.  This behavior can be understood
qualitatively in a model treating the excited electron-hole pairs
as a plasma.  
\end{abstract}   
   
\maketitle   
   
\section{Introduction} 
The Maxwell equations describe propagation of electromagnetic
fields in bulk matter taking into account the material properties by
the constitutive relations.
For ordinary light pulses, the response of the medium is linear in the
electromagnetic field and is characterized by the linear susceptibilities.
In recent experiments with intense and ultrashort laser pulses, 
however, one often encounter conditions which require theoretical 
treatments beyond the linear response. If the perturbative expansion
is no longer useful, one need to go back to the time-dependent 
Schr\"odinger equation for electrons and to solve it in time domain.

In the last two decades, computational approaches to solve the
time-dependent Schr\"odinger equation under intense
electric fields have been developed for atoms and
small molecules \cite{kr92,ch92,ka99}.
For electron dynamics in bulk matter as well as
in molecules, one often needs to go to the less demanding
approach based on time-dependent density-functional theory
(TDDFT) \cite{pe99,ca00,to01,no04,ca04,ot08,sh10,mi10}. 
We consider the TDDFT is 
the only {\it ab-initio} quantum method applicable to high fields 
in condensed media.

In this paper, we develop a formalism and computational method 
to describe propagation of intense electromagnetic field in the 
condensed medium incorporating feedback of electron 
dynamics to the electromagnetic field. This requires a consistent 
treatment of electrons and the electromagnetic field in coupled
equations of motion.  Such attempts have
been undertaken by several groups, for isolated molecules \cite{fr11}, 
nano-particles \cite{iw09,iw10,ch10}, and gases \cite{lo07}.
 
Experimentally, electron-hole plasmas are generated by irradiating
solids with strong laser pulses, and the threshold for dielectric breakdown has been measured 
\cite{re92,vo96,le98,so00,hu01,na02,ma04,ra05,da06,wi06}. 
To describe the phenomena, model approaches such as a rate
equation for electronic excitations have been developed
\cite{re04,pe05,re06,ha07,pe08,me10}.
For this problem, we have developed a first-principles approach 
\cite{ot08,ot09}. We calculated dielectric breakdown in crystalline
diamond \cite{ot08} and quartz \cite{ot09},  using TDDFT and treating the
electric field as a longitudinal field.  The calculated dielectric
breakdown threshold was much higher than observed. 
In the present work, we improve the theory by incorporating both 
magnetic and electric fields  in the equations,  permitting a proper 
description of transverse electromagnetic wave propagation.

Our formal development gives a way to separate out the two 
spatial scales that must be treated simultaneously.
The electron dynamics is calculated on the atomic scale,  
resolving position dependences of some tenths of an  
atomic unit. The electronic field is decomposed into two parts, 
one on the atomic scale and the other on the scale of the 
electromagnetic wave length. The atomic scale field is very 
similar from one unit cell to neighboring cells of the crystal. 
The other part of the field gives the large scale variation 
needed to describing the electromagnetic self-coupling and 
wave propagation. We introduce two grid systems with different 
resolution for this problem.  

The construction of the paper is as follows. In Sec. 2, we
present our formalism of multi-scale description for coupled
dynamics of electrons and electromagnetic fields.
In Sec. 3, numerical methods are explained.
In Sec. 4, calculated results are presented. We provide an
interpretation for the electron dynamics at the surface in
terms of the electron-hole plasma. We also compare dynamics
of the present multi-scale calculation with the microscopic 
dynamics in longitudinal and transverse geometries. Finally, 
a summary is presented in Sec. 5.

\section{Formalism}   
  
\subsection{Macroscopic equations for electromagnetic fields}   
 We will consider a coupled dynamics of electrons and electromagnetic   
field in bulk crystalline solid allowing for strong electromagnetic fields.
We immediately recognize there are two different spatial scales in  
the problem. The spatial scale of electromagnetic field is set by  
laser wavelength, of the order 1 $\mu$m. The spatial scale of electron   
dynamics is much smaller, of the order of $10^{-1} $ nm.  
We are thus led to a multi-scale description for the problem,  
employing two spatial grids of different grid sizes. We will use the
notation  $\mathbf R$ and $\vec r$ for the macroscopic and 
microscopic coordinates,  respectively.  

The essence of our method is to use the freedom to choose the 
electromagnetic gauge to separate the two scales.  In the expression
for the electric field, 
\be
\vec E = -\vec \nabla \phi - {1\over c}{\partial \vec A \over \partial t}
\ee
the gauge field $\vec A$ contains all the macroscopic electromagnetic
physics.  The microscopic physics, to be calculated on a unit cell
of the lattice, uses both $\vec A$ and the scalar potential $\phi$.   

To derive the theory formally, we start by taking a specific gauge
condition, the scalar potential $\phi$ is set equal to zero. 
In this gauge, we have the following equations for the vector potential
$\vec A(\vec r,t)$,
\be
-\frac{1}{c} \frac{\partial}{\partial t} \vec\nabla \vec A
= 4 \pi e \left(n_{ion}-n \right),
\label{lA}
\ee
\be
\frac{1}{c^2} \frac{\partial^2 \vec A}{\partial t^2}
-\nabla^2 \vec A 
+ \vec\nabla \left( \vec\nabla \cdot \vec A \right)
= - \frac{4\pi e}{c} \vec j.
\label{ltA}
\ee
Here we introduced the ionic density given by
\be
n_{ion}(\vec r) = \sum_{\alpha} Z_{\alpha} \delta(\vec r - \vec R_{\alpha}),
\ee
where $\vec R_{\alpha}$ and $Z_{\alpha}$ are the coordinate and charge
number of $\alpha$-th ion, respectively. We ignore the motion of ions
throughout this paper. $n$ and $\vec j$ are the number density and 
current of electrons, respectively, and satisfy the equation of continuity,
\be
\frac{\partial}{\partial t} n + \vec \nabla \vec j = 0.
\ee
We note that longitudinal part of the vector potential is described
redundantly by two equations (\ref{lA}) and (\ref{ltA}).

We then proceed to define macroscopic quantities from the microscopic
ones. As will be discussed later, the microscopic density and current
are obtained from the time-dependent Kohn-Sham orbitals.  
The macroscopic version of these quantities, $N_{\mathbf R}(t)$ and 
$\vec J_{\mathbf R} (t)$, may be defined in principle by applying some 
smoothing function to the microscopic quantities. In practice, we 
define this by averaging over the unit cell of the lattice.  
The macroscopic density and current also satisfies the equation
of continuity. We will not need it for the geometry considered below,
but it would be needed for other geometries.

We obtain macroscopic vector potential from $\vec A(\vec r,t)$
by a smoothing procedure which we denote as $\vec A_{\mathbf R}(t)$.
It satisfies the equations
\be
-\frac{1}{c} \frac{\partial}{\partial t} 
\vec\nabla_{\mathbf R} \vec A_{\mathbf R}(t) 
=
-4\pi e N_{\mathbf R}(t),
\label{macrolongA}
\ee
\be
\frac{1}{c^2} \frac{\partial^2}{\partial t^2} \vec A_{\mathbf R}(t)  
- \vec\nabla^2_{\mathbf R} \vec A_{\mathbf R}(t) 
+ \vec \nabla_{\mathbf R} \left(\vec \nabla_{\mathbf R} \cdot
\vec A_{\mathbf R}(t)\right)
= -\frac{4\pi e}{c} \vec J_{\mathbf R}(t),  
\ee
This is our basic equation to describe a propagation of macroscopic
electromagnetic field. 

For a microscopic physics, we treat the macroscopic field as uniform and
otherwise
we retain only a longitudinal part of the vector potential. 
Physically, the approximation is to neglect the transverse current
and variation of the magnetic field within the unit cells.
The neglect of transverse current amounts to ignore the orbital
magnetization. The neglect of magnetic field effects on electrons
may be justified when the velocity of electrons accelerated by the 
laser pulse is much smaller than the velocity of light.  
As will be explained later, we will employ a periodic scalar potential 
instead of the longitudinal vector potential for the microscopic 
description in the unit cells.

In general, the presence of boundaries requires special attention.  
If we may assume that a surface charge is localized in a sufficiently
thin layer at a surface, we may treat it as a discontinuity of the
macroscopic vector potential at the surface. Let us consider a
small volume around a point ${\mathbf R}$ at a surface and apply
the Gauss theorem to Eq.~(\ref{macrolongA}). We obtain
\be
\vec n \cdot \left( 
\vec A_{out,\mathbf R} - \vec A_{in,\mathbf R} \right)
=
4\pi c e \int^t dt' \Sigma_{\mathbf R}(t'),
\ee
where $\vec n$ is a unit vector normal to the surface at ${\mathbf R}$
and $\Sigma_{\mathbf R}(t)$ is the surface charge at ${\mathbf R}$.
This equation describes the boundary condition for the macroscopic
vector potential across the surface.
In the geometry we consider here, however, the fields are all parallel 
to the surface so that the macroscopic vector potential is continuous
at the surface.
  
\subsection{Microscopic equations for electrons}  
  
For the microscopic electron dynamics, we
assume a periodic band structure and apply the equations of 
motion of the time-dependent density functional theory.  
We will make several assumptions here.  

First we assume that electron dynamics at different   
macroscopic positions may be described independently.   
Namely, we define Kohn-Sham orbitals   
at every macroscopic grid point and ignore any direct   
interactions between electrons belonging to different   
macroscopic grid points. We only take into account the  
interaction between electrons of different macroscopic   
grid points through the macroscopic vector potential 
$\vec A_{\mathbf R}(t)$.

Second, we assume that $N_{\mathbf R}(t)$ is independent of time.
This condition is satisfied in the one-dimensional  
propagation of linearly polarized light at normal incidence on 
an interface, since the macroscopic current does not include
any longitudinal component as discussed below.
The orbitals evolve under the time-dependent
Kohn-Sham equations, but the number of electrons in each cell
remains the same, and in fact the orbital occupation numbers
remain zero or one in the time-evolved basis.

Third, within each cell of the microscopic scale, we ignore 
any effects of magnetic fields on the electrons.
The macroscopic vector potential $\vec A_{\mathbf R}(t)$
will be treated as a uniform field in the microscopic scale.  
This permits us to treat electron dynamics induced by a
uniform electric field. We also ignore the transverse 
component of the microscopic vector potential, retaining 
only the longitudinal part as mentioned before.

Since all that matters are the physical fields, we are permitted
to make a different choice of gauge for the microscopic fields.
In Eqs.~(\ref{lA}) and (\ref{ltA}), we had chosen the gauge condition
that removes the scalar potential. 
However, to take advantage of the periodicity of the lattice, 
we make a gauge transformation at each macroscopic grid point, 
expressing periodic electromagnetic field with a scalar potential 
$\phi$ instead of the longitudinal part of the microscopic vector 
potential. We denote the scalar potential at macroscopic grid point
${\mathbf R}$ as $\phi_{\mathbf R}(\vec r,t)$, to indicate 
that $\mathbf R$ will just be a parameter in the equation 
of motion for $\phi$. 

We denote the Kohn-Sham orbitals at a macroscopic coordinate  
$\mathbf R$ as $\psi_{i,\mathbf R}(\vec r,t)$. Under above   
conditions and assumptions, the time-dependent Kohn-Sham  
(TDKS) equation may be written as 
\begin{equation}  
i\hbar \frac{\partial}{\partial t} \psi_{i,\mathbf R}  (\vec r,t)
= \left\{ \frac{1}{2m} \left( -i\hbar \nabla_{\vec r}  
+ \frac{e}{c} \vec A_{\mathbf R}(t) \right)^2
-e \phi_{\mathbf R}(\vec r, t) 
+ \frac{\delta E_{xc}}{\delta n} \right\} \psi_{i,\mathbf R}(\vec r,t). 
\label{TDKS}  
\end{equation}  
In solving Eq.~(\ref{TDKS}), the macroscopic coordinate   
$\mathbf R$ is treated as a parameter. 
The Kohn-Sham Hamiltonian thus defined
is periodic in space and one may introduce Bloch functions at   
each time step, applying periodic boundary conditions on the 
electron orbitals within each microscopic cell\cite{be00,ot08,sh10}.  

The electron density and current are both periodic in space  
and are given by  
\begin{equation}  
n_{\mathbf R}(\vec r,t)=  
\sum_i  \vert \psi_{i,\mathbf R}(\vec r,t) \vert^2,
\end{equation}  
\begin{eqnarray}  
\vec j_{\mathbf R}(\vec r,t)  
=  \frac{1}{2m} \sum_i  
&&  \left\{ \psi^*_{i,\mathbf R}(\vec r,t)   
\left( -i\hbar \vec\nabla_{\vec r}  
	+\frac{e}{c}\vec A_{\mathbf R}(t) \right)  
\psi_{i,\mathbf R}(\vec r,t)  
\right. \nonumber\\  
&& \left. -   
\psi_{i,\mathbf R}(\vec r,t)  
\left( i\hbar \vec\nabla_{\vec r}  
+\frac{e}{c}\vec A_{\mathbf R}(t) \right)  
\psi^*_{i,\mathbf R}(\vec r,t)  
\right\},
\end{eqnarray}
where the sum $i$ is over occupied orbitals.  
The scalar potential $\phi_{\mathbf R}(\vec r,t)$ satisfies  
the Poisson equation,  
\begin{equation}  
\nabla^2_{\vec r} \phi_{\mathbf R}(\vec r,t)  
= -4\pi \left( e n_{ion,\mathbf R}(\vec r) - e n_{\mathbf R}(\vec r,t) \right),  
\end{equation}  
where $n_{ion,\mathbf R}$ is the ionic density at macroscopic grid point
${\mathbf R}$. 
  
Since the density and the current are periodic in space,    
the average over the unit cell is meaningful also on the macroscopic 
scale.  The main macroscopic quantity we need from the electronic
dynamics is the current, defined as
\begin{equation}  
\vec J_{\mathbf R}(t) = \frac{1}{\Omega} \int_{\Omega} d\vec r  
\vec j_{\mathbf R}(\vec r,t),
\end{equation}  
where $\Omega$ is the volume of the unit cell.  

\subsection{Conserved energy}   
  
To obtain an expression for the conserved energy in the present  
multi-scale description, we first note the above equations of 
motion may be derived from the following Lagrangian.  
\begin{eqnarray}  
L=\int d{\mathbf R} &&  
\left[ \sum_i \int_{\Omega} d\vec r  
\left\{ \psi^*_{i,\mathbf R}   
i\hbar \frac{\partial}{\partial t} \psi_{i,\mathbf R}  
-\frac{1}{2m} \left\vert \left( -i\hbar \vec\nabla_{\vec r}  
+\frac{e}{c}\vec A_{\mathbf R} \right) \psi_{i,\mathbf R}  
\right\vert^2 \right\} \right.  
\nonumber\\  
&&   
-\int_{\Omega}d\vec r \left\{ \left(  
en_{ion}-en_{\mathbf R} \right) \phi_{\mathbf R}   
-E_{xc}[n_{\mathbf R}] \right\}   
\nonumber\\  
&&  
\left. +\int_{\Omega} d\vec r \frac{1}{8\pi}  
\left( \vec\nabla_{\vec r} \phi_{\mathbf R} \right)^2  
+ \frac{\Omega}{8\pi c^2}   
\left( \frac{\partial \vec A_{\mathbf R}}{\partial t} \right)^2  
- \frac{\Omega}{8\pi}\left( \vec\nabla_{\mathbf R} \times  
\vec A_{\mathbf R} \right)^2 \right].  
\end{eqnarray}  
From this Lagrangian, one may derive the following Hamiltonian,  
\begin{eqnarray}  
H=\int d\vec R && \left[  
\sum_i  \int_{\Omega} d\vec r  
\frac{1}{2m} \left\vert \left( -i\hbar \vec\nabla_{\vec r}  
+\frac{e}{c}\vec A_{\mathbf R} \right) \psi_{i,\mathbf R}  
\right\vert^2 \right.  
\nonumber\\  
&& \left.  
+\int_{\Omega}d\vec r \left\{ \frac{1}{2} \left(  
en_{ion}-en_{\mathbf R} \right) \phi_{\mathbf R}   
+E_{xc}[n_{\mathbf R}] \right\} \right.  
\nonumber\\  
&&  
\left. + \frac{\Omega}{8\pi c^2}   
\left( \frac{\partial \vec A_{\mathbf R}}{\partial t} \right)^2  
+ \frac{\Omega}{8\pi}\left( \vec\nabla_{\mathbf R} \times  
\vec A_{\mathbf R} \right)^2 \right].  
\end{eqnarray}  
The energy calculated from this Hamiltonian is conserved by
the equations of motion.
  
\subsection{One-dimensional propagation}  
  
In this paper, we will consider a propagation of linearly  
polarized laser pulse incident normally on a bulk   
crystalline Si on the  [110] surface with the laser electric   
field in [100] direction. There 
are two spatial regions on the macroscopic scale, vacuum and crystalline solid.   
We take a macroscopic coordinate system such that the   
surface of the crystalline solid is $xy$-plane with $z=0$.   
In this geometry, macroscopic quantities are uniform in   
both $x$- and $y$-directions. Therefore, macroscopic   
quantities are specified by the $z$-component of $\mathbf R$   
which we denote ${\mathrm Z}$.  
 
The macroscopic vector potential has the following form,  
\begin{equation}  
\vec A_{\mathbf R}(t) = \hat x A_{\mathrm Z}(t),  
\end{equation}  
In the vacuum region $({\mathrm Z} <0)$, $A_{\mathrm Z}(t)$ 
is composed of incident  and reflected waves. Inside the 
solid,   
we assume a locally dipole approximation at each macroscopic   
coordinate, as mentioned before. Then the macroscopic electron  
current is parallel to the vector potential.  
\begin{equation}  
\vec J_{\mathbf R}(t) = \hat x J_{\mathrm Z}(t).  
\end{equation}  
We note that this form of electric current is transverse.  
This justifies our assumption that the macroscopic electron
density $N_{\mathbf R}(t)$ is independent of time.

For later convenience, we summarize equations of motion in the  
one-dimensional geometry.  
The vector potential $A_{\mathrm Z}(t)$ satisfies the following equation,  
\begin{equation}  
\label{Aint}
\frac{1}{c^2} \frac{\partial^2}{\partial t^2} A_{\mathrm Z}(t)  
- \frac{\partial^2}{\partial Z^2} A_{\mathrm Z}(t)  
=  
-\frac{4\pi e}{c} J_{\mathrm Z}(t).  
\end{equation}  
The TDKS equation is given by  
\begin{equation}  
i\hbar \frac{\partial}{\partial t} \psi_{i,\mathrm Z}(\vec r,t)  
= \left\{ \frac{1}{2m} \left(  
-i\hbar \vec\nabla_{\vec r}+\frac{e}{c}\hat x A_{\mathrm Z}(t) \right)^2 
-e\phi_{\mathrm Z}(\vec r,t) + \frac{\delta E_{xc}}{\delta n} \right\}  
\psi_{i,\mathrm Z}(\vec r,t),  
\label{TDKSZ} 
\end{equation}  
with the density and current,  
\begin{equation}  
n_{\mathrm Z}(\vec r,t)=\sum_i \vert \psi_{i,\mathrm Z}(\vec r,t) \vert^2,  
\label{nZ} 
\end{equation}  
\begin{equation}  
\vec j_{\mathrm Z}(\vec r,t) = \frac{1}{2m} \sum_i 
\left\{ \psi^*_{i,\mathrm Z} \left( -i\hbar \vec\nabla_{\vec r}  
+\frac{e}{c}\hat x A_{\mathrm Z} \right) \psi_{i,\mathrm Z}  
- \psi_{i,\mathrm Z} \left( i\hbar \vec\nabla_{\vec r}   
+ \frac{e}{c}\hat x A_{\mathrm Z} \right) \psi^*_{i,\mathrm Z} \right\},  
\label{jZ} 
\end{equation}  
\begin{equation}  
J_{\mathrm Z}(t) = \frac{1}{\Omega} \int_{\Omega} d\vec r 
\hat x \vec j_{\mathrm Z}(\vec r,t).  
\label{macroJZ} 
\end{equation}  
The energy per unit area $E_{\cal A}$ is a conserved quantity and is
given by  
\begin{eqnarray}  
E_{\cal A}  = {1\over \Omega}\int d{\mathrm Z} 
&& \left[  
\sum_i \int_{\Omega} d\vec r  
\frac{1}{2m} \left\vert \left(  
-i\hbar \vec\nabla_{\vec r} + \frac{e}{c}\hat x A_{\mathrm Z} \right)  
\psi_{i,\mathrm Z} \right\vert^2 \right.  
\nonumber\\  
&&+\int_{\Omega} d\vec r   
\left\{ \frac{1}{2}  
(e n_{ion,\mathrm Z} -e n_{\mathrm Z})\phi_{\mathrm Z} 
+ E_{xc}[n_{\mathrm Z}] \right\}  
\nonumber\\  
&& \left. +\frac{\Omega}{8\pi }\left\{   \frac{1}{c}
\left( \frac{\partial A_{\mathrm Z}}{\partial t} \right)^2  
+ \left(  
\frac{\partial A_{\mathrm Z}}{\partial \mathrm Z} \right)^2 \right\}\right]\,.  
\end{eqnarray}  

\subsection{Linear response}  
 
It is essential that the theory properly describes the propagation
of electromagnetic waves  in the weak field limit, if it is to
be useful more generally.  In weak fields the electron dynamics can 
be calculated perturbatively to arrive at the usual linear response.
The present formalism gives the same linear response as other
approaches, so the weak field limit will be correct if the
dielectric function is given correctly.
In our previous work,
we calculated the linear response by separating $A$ into a
part that arose from external sources and a part arises from
the medium \cite{be00}.  It is not necessary to make this separation in 
the present formalism.  
To derive the dielectric function, we note 
Eqs.~(\ref{TDKSZ}), (\ref{jZ}), 
and (\ref{macroJZ}) describe relation between the macroscopic 
vector potential $A_{\mathrm Z}(t)$ and the macroscopic current 
$J_{\mathrm Z}(t)$.  We may summarize the relation as, 
\begin{equation} 
\label{A2J}
J_{\mathrm Z}(t) = \int^t dt' \sigma(t-t') E_{\mathrm Z}(t') 
= -\frac{1}{c} \int^t dt' \sigma(t-t') \frac{\partial A_{\mathrm Z}(t')}{\partial t}, 
\end{equation} 
where we have introduced the electric conductivity function
$\sigma(t)$.  In the microscopic TDDFT calculation, the vector potential
is an external variable.  We may compute $J_{\mathrm Z}(t)$ for an arbitrary 
$A_{\mathrm Z}(t)$ and thus determine the conductivity function.  Since the
equation is linear, it is easy to extract the frequency-dependent
conductivity $\sigma(\omega)$.  The dielectric function $\epsilon(\omega)$
is then 
given by the usual formula
\begin{equation} 
\epsilon(\omega) = 1 + \frac{4\pi i \sigma(\omega)}{\omega}. 
\label{eps-sgm}
\end{equation} 
Further details of calculating the dielectric function in this 
formalism are given in the Appendix.

Since our theory gives the same macroscopic current as in the linear 
response, the macroscopic equations only require the dielectric
function from the microscopic dynamics.  
Thus, the propagation of electromagnetic waves will be given by 
the usual relation between wave vector $k$ and frequency $\omega$,
\be
\omega = \frac{ck}{\sqrt{\epsilon(\omega)}}.
\ee

\section{Numerical}

\subsection{Units}

For numerical quantities on the microscopic scale, we will use
atomic units for length and field strengths.  However, we will
use micrometers for lengths on the macroscopic scale.  On both scales,
energies will be in eV units and time in femtosecond units. We
continue to include the dimensionful quantities $m$ and $e$ in
formulas even though their values are equal to one.  
For the laser intensity, we will use the conventional units of  W/cm$^2$.  
The conversion factor to atomic units is 
1 a.u. = $3.509 \times 10^{16}$ W/cm$^2$. 

\subsection{Electronic scale}  
As in our previous applications to crystalline materials\cite{be00,ot08}, 
we calculate the evolution of the electron wave function in a unit cell 
of the crystal.  The orbital wave functions are represented on a 3D 
spatial grid which typically has a dimension of $16^3$.  The Si lattice constant 
is 10.26 a.u. giving a mesh space of $\Delta x = 0.64$ a.u.  A high order
finite difference formula is used for the derivative calculations \cite{ch94}.
The  number of $k$-points in the reciprocal space cell is taken as 
$8^3$; however  due to symmetry there are only 80 distinct orbitals 
to be calculated.

The number of $k$-points adopted here are smaller than those
employed in our previous work \cite{ot08, sh10}. The present choice
is decided from a computational feasibility.
Our present scheme requires microscopic electron dynamics calculations 
in a number of macroscopic grid points simultaneously.
Thus the present calculation consumes much more computational 
resources than our previous calculations of a single microscopic
electron dynamics. The computational wall time with the above 
setting is approximately 15 hours employing 1,024 processor cores 
in parallel with Intel Xeon X5570 (2.93GHz). 
This is close to the limit of our computational capability
at present. The present calculation with $8^3$ $k$-points may not 
provide fully convergent results but we do not expect this truncation
to affect the physical results by more than 10 percent.

The electronic structure of each macroscopic position is initialized by the ground state
Kohn-Sham orbitals.
The  wave functions are evolved by using the 4-th order
expansion of the TDDFT evolution operator \cite{ya96,ya06}
\be  
e^{-i H \Delta t} \approx \sum_{n=0}^4 {(i \Delta t)^n\over n!} H^n,  
\label{Taylor4}
\ee  
where $H$ is the Kohn-Sham single-particle Hamiltonian appearing
in Eq.~(\ref{TDKSZ}). The orbitals at time $t+\Delta t$ are computed 
by applying Eq.~(\ref{Taylor4}) to the orbitals at time $t$.  
The Kohn-Sham operator $H$ is a function of the fields $\phi$, $A$,
and the density.  We use a fixed-time Hamiltonian $H$ in which 
the scalar potential and the density are taken at time $t$ and the 
field $A$ is taken to be $(A(t)+A(t+\Delta t))/2$.  
This prescription does not require any
predictor step (see below) and it gives good energy conservation 
over the course of the integration time.

The algorithm (\ref{Taylor4}) is stable provided $\Delta t$ satisfies 
the condition\cite{be06}
\be  
\Delta t < \sqrt{\frac{2}{9}}m (\Delta x)^2 \approx 0.2\,\,\,{\rm a.u.}
\ee  
We use a somewhat smaller value in the calculations below, 
$\Delta t = 0.08$ au.  
In a typical run, the equations of motion are integrated for 16,000
time steps, amount to a total time $1280$ au  $= 31$ fs.  This is sufficient
to see the passage of a femtosecond laser pulse through a section of the
solid.  At longer times, other processes such as thermalizing collisions
and ionic motion become important, and the TDDFT dynamics is no longer
valid.
  
As in previous work, we use the adiabatic approximation, taking the
time-dependent functional in TDDFT the same as the ground state
functional.  We use the functional of Ref. \cite{lda} in the local
density approximation.
The electron-ion interaction is treated with a norm-conserving
pseudopotential \cite{tm} with a gauge correction for the nonlocal
part \cite{be00}.

\subsection{Macroscopic Scale}

On the macroscopic scale, $A_{\mathrm Z}(t)$ and $J_{\mathrm Z}(t)$ 
are considered as continuous functions, but they are discretized for 
the numerical calculation.  In the results presented in the next section, 
we use a mesh size of 250 au $\simeq$ 13 nm. This permits us to propagate
the pulse over a distance of several $\mu$m in the medium sampling 
the microscopic dynamics at several hundred points. We employ 256
grid points. The integrator for Eq. (\ref{Aint}) is straightforward, 
but the coupling between scales requires some care.  
The following update procedure is simple and 
conserves energy to adequate precision:
\be
A_{\mathrm Z} (t+\Delta t) := 
2 A_{\mathrm Z} (t)- A_{\mathrm Z} (t-\Delta t)
+ c^2 \Delta t^2 \left\{
{d^2\over d {\mathrm Z}^2} A_{\mathrm Z} (t) 
+{4\pi e^2\over c} J_{\mathrm Z}(t) \right\},
\ee
where the space derivative is treated with a simple three-point
formula. It also permits us to use $A_{\mathrm Z}(t+\Delta t)$ 
when updating the variables for the microscopic scale.

\subsection{Laser field}

We use the following functional form for the shape of the incident
laser pulse,
\be
E_{\mathrm Z}(t) = E_0 
\sin^2({\pi ({\mathrm Z}-ct-{\mathrm Z}_0)\over cT}) 
\sin({\omega_\ell ({\mathrm Z}-ct-{\mathrm Z}_0) \over c}),
\hspace{10mm} ({\mathrm Z}_0 < {\mathrm Z}-ct < {\mathrm Z}_0+cT).
\ee
Here $T= 18$ fs controls the pulse width, 
$\omega_\ell=1.55$ eV/$\hbar$ is the laser frequency, and 
$E_0$ is the maximum electric field strength which is related
to the laser intensity $I_0$ by $I_0=cE_0^2/8\pi$. 

The corresponding gauge field is obtained by an analytic
calculation of the following integral,
\be
A_{\mathrm Z}(t) = -c \int^t_{-\infty} dt' E_{\mathrm Z}(t').
\ee
To start calculation, we need the initial vector potential at two 
times. One is given by $A_{\mathrm Z}(t=0)$. Instead of using analytic 
form, we employ the following for the other,
\be
A_{\mathrm Z}(\Delta t) = A_{\mathrm Z}(0) 
+ \Delta t \frac{\partial A_{\mathrm Z}}{\partial t}(0)
+ \frac{1}{2} c^2 \Delta t^2 \frac{\partial^2 A_{\mathrm Z}}
{\partial {\mathrm Z}^2}(0).
\ee

\section{Results}  

\subsection{Pulse propagation}
\begin{figure}  
\includegraphics [width = 11cm]{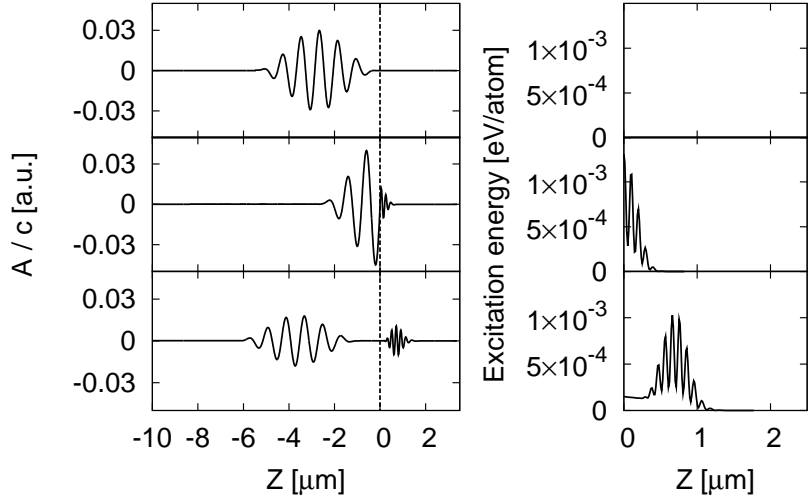}  
\caption{\label{t-dependence} Snapshots of the electromagnetic  
fields  (vector potential divided by light speed, $A/c$, left panels) 
and of the electronic excitation energy per 
atom (right panels) at different times, shown as a function of macroscopic 
position. The vacuum is at ${\mathrm Z}<0$ and the Si crystal
is at ${\mathrm Z}>0$.  Top panels:  initial starting field, with 
pulse on left moving toward the Si surface.  Middle panel:  
at the point the middle of the pulse reaches the surface.  
Lower panels:  the reflected and transmitted pulses are well 
separated. The maximum intensity of the incident laser pulse is set
$10^{11}$ W/cm$^2$.
}  
\end{figure}  
We first note that the calculated dielectric constant at the laser
frequency, $\epsilon(\omega_{\ell}) = 16.2$,  is 
in reasonable agreement
with the observed value, $\epsilon(\omega_\ell)=13.6$.  
See the Appendix for details of the calculated dielectric function.  The
most significant shortcoming of the TDDFT dielectric function is this
too-small band gap.  Apart from that, we can confident that 
the present calculations will be reliable in the weak field limit.

Snapshots of the time evolution for a typical run are shown in 
Fig.~\ref{t-dependence}.  The initial laser pulse at $t=0$ has a 
peak intensity of $I_0= 10^{11}$ W/cm$^2$ and started at a 
position ${\mathrm Z}= -2.9~\mu$m with respect to the Si 
surface at ${\mathrm Z}=0$. This is shown in the upper panel 
of the figure.  
The middle panel shows the field when the center of the pulse 
has just reached the surface, at $t=9.6$ fs.  One can see a 
transmitted wave of much smaller amplitude.  
In the lower panel, at $t=21.3$ fs, the wave has
completely separated into the reflected and transmitted components.
The wave length of the transmitted component can be read off as
$\lambda_m = 3770$ au, consistent with the low-field formula
$\lambda_m = \lambda/\sqrt{\varepsilon}\approx 3800$ au.  
The center of the transmitted pulse is at ${\mathrm Z}=0.71~ \mu$m.  
Taking the propagation
time from the surface to be $t_2-t_1$, the wave speed from the
calculation is $0.20c$.  This is somewhat less than the phase velocity, 
which is $c/\sqrt{\varepsilon}\approx 0.25c$, but is completely 
consistent with the low-field group velocity computed as
\be
v_g = {c \over \sqrt{\varepsilon}
\left( 1 + {\omega\over 2\varepsilon}{d \varepsilon \over d \omega}\right)}.
\ee
We also observe a chirp effect on the transmitted wave, 
stretched out at the front and condensed at the end of the
transmitted pulse.

We next examine the reflected and transmitted intensities. 
The maximum amplitude in Fig. \ref{t-dependence} for the initial 
pulse is $A_0/c=0.0298$, for the reflected pulse is $A_r/c=0.0180$, 
and for the transmitted pulse is $A_t/c=0.0107$.  We obtain for 
the calculated reflectivity $r\equiv (A_r/A_0)^2\approx 0.36$.
The reflectivity according to dielectric theory is given by
\be
\label{R}
R = \left|{\sqrt{\epsilon}-1\over \sqrt{\epsilon}+1}\right|^2
\ee
at normal incidence.  With our theoretical value for 
$\epsilon(\omega_\ell)$, we obtain $R=0.36$, in good agreement 
with the real-time dynamics.   
The transmitted intensity is more complicated, since there are 
contributions from both the electronic part and the field part
and the wave velocity is different.
We can still ask how well the observed field amplitude agrees 
with dielectric theory.  Expressing the transmittance $T$ in terms
of the field amplitudes, the formula is
\be
T = \sqrt{\epsilon}\left({A_t\over A_0}\right)^2 .
\ee
This gives $T=0.52$ for the case shown in Fig.~\ref{t-dependence}.  
The dielectric transmittance can also be expressed purely 
in terms of $\epsilon$, giving $T=1-R\approx 0.64$.  The
difference between the two numbers, $0.64-0.52$, is due to absorption.
Thus the theory predicts that 12\% of the energy is absorbed
in the first 20 fs for an pulse of strength $10^{11}$ W/cm$^2$. 
In fact, as may be seen in the bottom right panel of Fig.~\ref{t-dependence}, we find a certain
fraction of the excitation energy is left in the spatial region where 
the laser pulse already passed. 
Notice that this  energy loss is not evident from the reflectance, 
which is still consistent with dielectric theory.

\begin{figure}  
\includegraphics [width = 11cm]{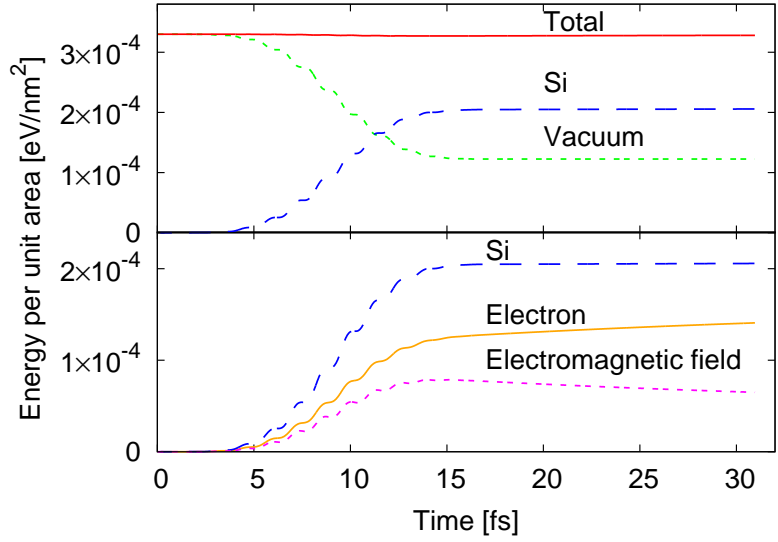}  
\caption{\label{e-cons} 
Energies per unit area integrated over macroscopic
coordinate ${\mathrm Z}$ are shown as a function of time. 
In the upper panel, the energies integrated
over ${\mathrm Z}<0$ (vacuum), ${\mathrm Z}>0$ (Si), and the whole
region (Total) are compared. In the lower panel, the energy integrated
over Si crystal region is decomposed into the field part and
the electronic excitation part. The incident laser pulse is the same
as that of Fig.~\ref{t-dependence}.
}  
\end{figure}  
In Fig.~\ref{e-cons}, we show energies per unit area integrated over 
the macroscopic coordinate. In the upper panel, the energy is
decomposed into vacuum region (${\mathrm Z}<0$, green dotted line) 
and Si crystal region (${\mathrm Z}>0$, blue dashed line). The sum
of the two contributions is shown by red solid line, showing that
the total energy is well conserved during the whole period.

In the lower panel, the energy per unit area in the Si crystal region
is decomposed into contributions of electronic excitations and
electromagnetic fields. Since the electromagnetic fields are separated
into reflected and transmitted fields after 15 fs, the energy of
Si crystal region does not change in that period. The energy of
transmitted electromagnetic fields decreases gradually as it is transferred to electronic excitation.

\begin{figure}  
\includegraphics [width = 11cm]{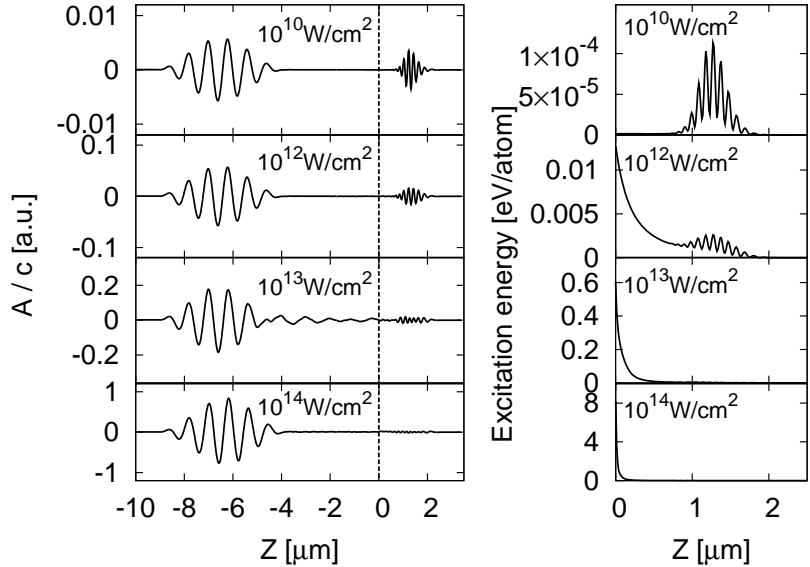}  
\caption{\label{I-dependence} 
State of the system at $t=21$ fs after the peak of the incident  
pulse reaches the surface for several different intensities of the
incident laser pulse. Left: the field divided by light speed, $A/c$; 
right: excitation energy per atom in the Si crystal.  
}  
\end{figure}  
We next show reflected and transmitted electromagnetic fields
at different intensity levels.
In the left panels of Fig.~\ref{I-dependence}, the vector potentials
are shown at a time when the transmitted and reflected waves 
are well separated. In the right panels, the electronic excitation
energies per atom is shown in the Si crystal region.
At the lowest intensity, the propagation of electromagnetic 
fields are well described by dielectric response. Essentially all of
the energy remains associated with the propagating transmitted
pulse. As the incident intensity increases, the transmitted wave 
becomes weaker than that expected from the linear response. 
We also find the central part of the transmitted pulse is suppressed 
strongly, producing a flat envelope of the pulse.
In contrast, the envelope of the reflected wave does not
change much in shape even at the highest intensity.
We also find, at the intensity of $10^{13}$ W/cm$^2$, an emission
of electromagnetic field is seen from the surface following the
main pulse of reflected wave. 
From the right panels, above $10^{12}$ W/cm$^2$ one sees 
that most of the energy is deposited in the medium with just 
a small fraction remaining in the transmitted electromagnetic pulse.
The deposition rate falls off with depth as to be expected from
the weakening of the pulse. At higher intensities the absorption rate greatly 
increases.  At $I_0=10^{13}$ W/cm$^2$ and higher the transmitted
pulse is almost completely absorbed in the first tenths of a $\mu$m.

\begin{figure}  
\includegraphics [width = 11cm]{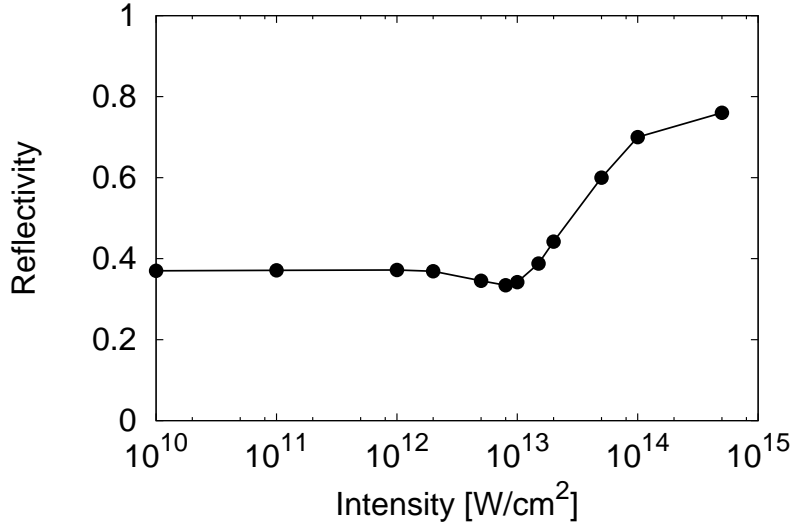} 
\caption{\label{r} 
The reflectivity of Si at normal incidence is shown as a function of peak 
laser intensity.
}  
\end{figure}  
In Fig.~\ref{r}, we show the reflectivity as a function of incident
laser intensity. Below $10^{12}$ W/cm$^2$, the reflectivity is
constant and in accord with dielectric theory (Eq. (\ref{R})).
Above $10^{12}$ W/cm$^2$, the reflectivity dips slightly,
showing a minimum around $10^{13}$ W/cm$^2$.
Above that intensity, the reflectivity start to increase gradually and
finally reach 0.75 at the intensity of $5 \times 10^{14}$ W/cm$^2$.
This behavior of reflectivity qualitatively follows the observed
evolution with intensity  \cite{so00}, where it was interpreted in a 
dielectric model including effects of the excited electrons.
We will later compare this model with our calculated reflectivity
function.

\subsection{Excitation in surface layer}
\label{surface-layer}
\begin{figure}  
\includegraphics [width = 11cm]{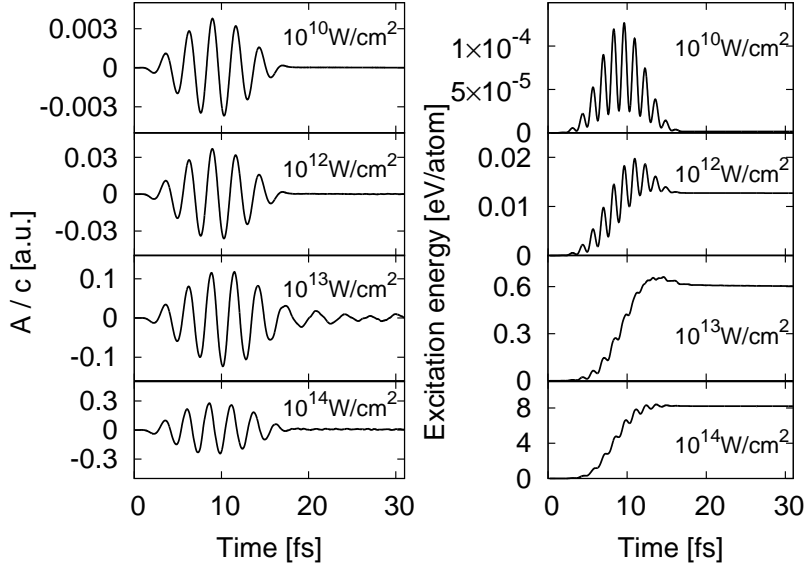} 
\caption{\label{t-surface} 
The vector potential divided by light speed (left panels)
and electronic excitation energy (right panels) at the
surface cell are shown as a function of time.
}  
\end{figure}  

We next examine in more detail the first cell at the surface.
The left-hand panel of Fig.~\ref{t-surface} shows the vector potential  
as a function of time for several laser intensities. 
From a dielectric response, we expect the field 
inside the Si crystal is related to the incident field by
$A_t = (2/1+\sqrt{\epsilon}) A_i$.
This relation holds well below $10^{12}$ W/cm$^2$.
At higher intensities, the field is less than this estimate gives.
We also observe an oscillation of the vector potential after 
the incident pulse ends at $10^{13}$ W/cm$^2$, in accordance
with what we found in Fig.~\ref{I-dependence}.
We will later consider this phenomenon with a model
dielectric function.

The electronic excitation energy in the first cell is shown in
the right-hand panel of Fig. \ref{t-surface}. At the lowest intensities,
the electronic energy is carried by the transmitted wave and leaves the
cell after passage of the pulse.
As the laser intensity increases, energy is transferred irreversibly
to electronic excitation, and reaches a plateau after the laser pulse
has passed  ($t>15$ fs).
This is because the only mechanism to transfer energy between macroscopic
grid points is through the macroscopic electromagnetic fields.
\begin{figure}  
\includegraphics [width = 8cm]{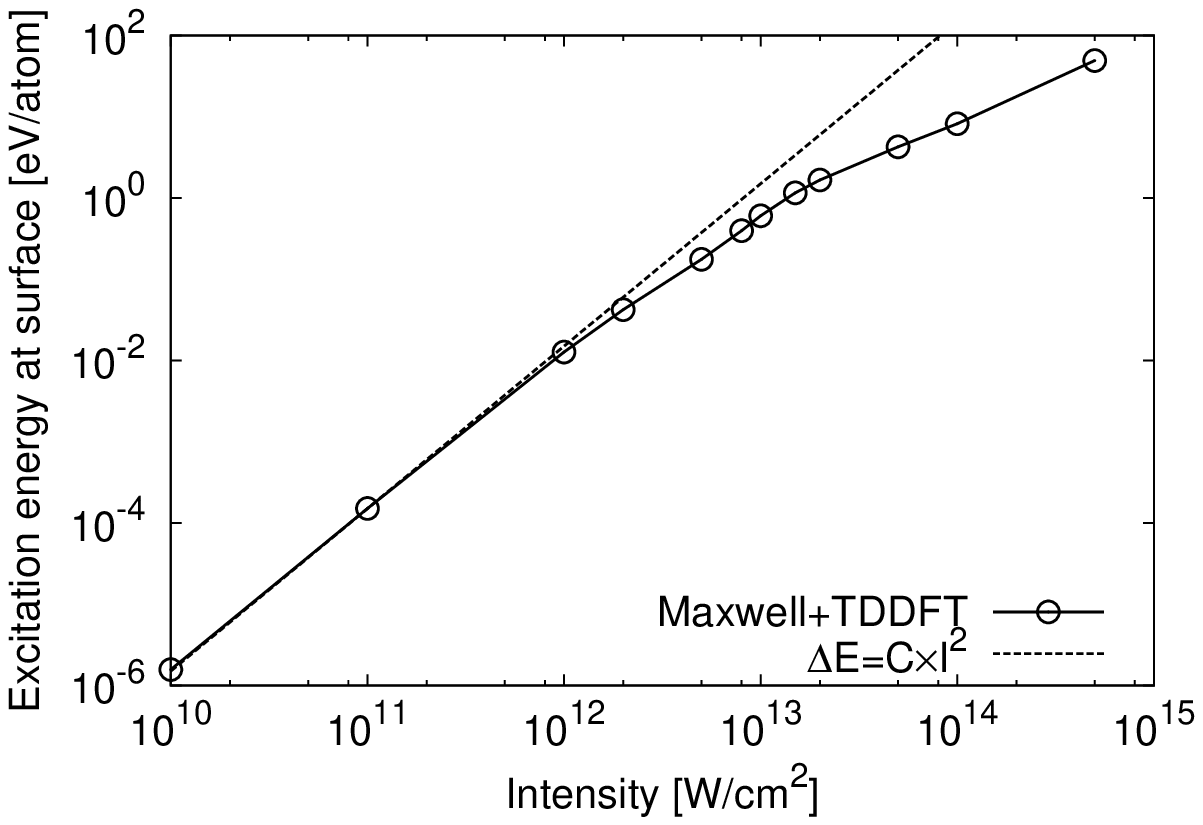} 
\includegraphics [width = 8cm]{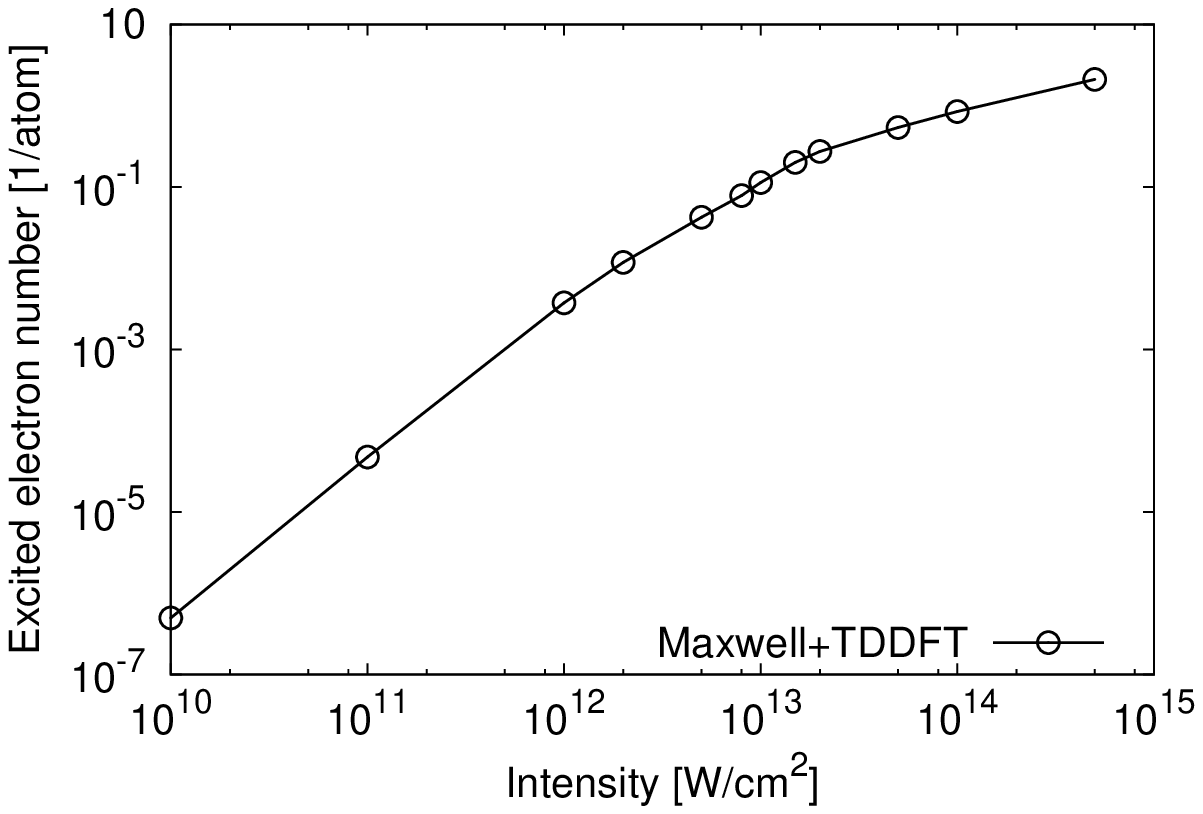}  
\caption{\label{E-surface} 
Excitation of the first layer of Si after the laser pulse ends.
Excitation energy per Si atom (left panel) and density of 
electron-hole pairs as number of pair per Silicon atom
(right panel) are shown as a function of laser intensity.
}  
\end{figure}  

Figure~\ref{E-surface} shows some final-state properties of the surface as a 
function of intensity.  The residual excitation energy is shown in the
left-hand panel.
At low intensities, the energy deposited is proportional to $I_0^2$.  
This is the expected dependence for  two-photon absorption.  
This is the most favorable absorption process in view of the photon 
energy: single-photon absorption is forbidden below the direct 
band gap, but the two-photon process is allowed.  
At $I_0\approx10^{13}$ W/cm$^2$ the excitation energy is 
0.6 eV per Si atom.  This energy is in the form of electron-hole pairs.   
The minimum energy of a pair is at the direct band gap,  2.4 eV.  
However, the excitation process forms a coherent pair with energies
distributed across the valence and conduction bands.
In the TDDFT dynamics, the coherence is lost after the pulse moves on, 
but the energy distribution remains the same.  

The number of particle-hole pairs $n_{ph}$ in the cell does not change after
the electromagnetic field has passed.  Then the number be calculated as the 
sum of overlaps of the time-dependent orbitals and the original 
Kohn-Sham orbitals,
\be
n_{ph} =
\sum_i \left\{ 1 - \sum_j  \vert \langle \psi_{j,{\mathrm Z}}(0) \vert
\psi_{i,{\mathrm Z}}(t) \rangle \vert^2 \right\},
\ee
where the sum over $i,j$ is taken over occupied orbitals. 
The results are shown in the right panel of Fig. \ref{E-surface}.  
\begin{figure}  
\includegraphics [width = 11cm]{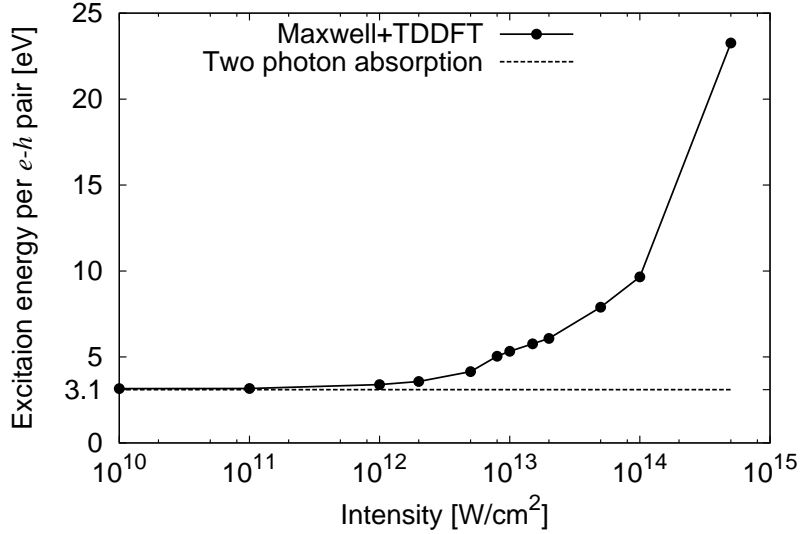} 
\caption{\label{Eperpair}
Electronic excitation energy per electron-hole pair as a
function of laser intensity.
}  
\end{figure}  
As seen from the figure, the density increases quadratically
with $I_0$ up to to a point and then continues to increase 
more gradually.  The ratio of energy density to particle-hole
pair density, shown in Fig.~\ref{Eperpair}, has a simple 
interpretation.  At low intensities, up to about $10^{12}$ W/cm$^2$,
it coincides 
accurately with two-photon
energy $2\hbar \omega_\ell = 3.1$ eV. 
The energy per pair gradually 
increases at higher intensities.  There
one may expect two processes which increase the
energy per pair. One is higher-order multiphoton absorption, as has 
been often discussed \cite{ke65,re80}. The other is the secondary 
excitation of electrons which have already been excited.  

With the information about the particle-hole density $n_{ph}$, 
we may interpret the reflectivity curve (Fig. 4) with a 
model for the dielectric function that includes plasma effects.
For example, in Ref. \cite{so00} and \cite{me10}, the response
of electrons excited in conduction band is described with
the Drude model. We consider the following simplified form for
the dielectric function,
\be
\label{eps-model}
\varepsilon(\omega,n_{ph}) = \varepsilon(\omega,0)
- \frac{4 \pi e^2 n_{ph}}{m^*}
\frac{1}{\omega \left( \omega + \frac{i}{\tau} \right) }\,.
\ee
Here $\varepsilon(\omega,0)$ is the dielectric function in the ground 
state; $m^*$ and $\tau$ are parameters of the Drude model.  For our
comparison we take $\varepsilon(\omega,0)$ from the linear response
(see Appendix) at $\omega=\omega_\ell= 1.55$ eV.  The reflectivity 
associated with the model dielectric function is determined from
Eq. (\ref{R}).  
\begin{figure}  
\includegraphics [width = 11cm]{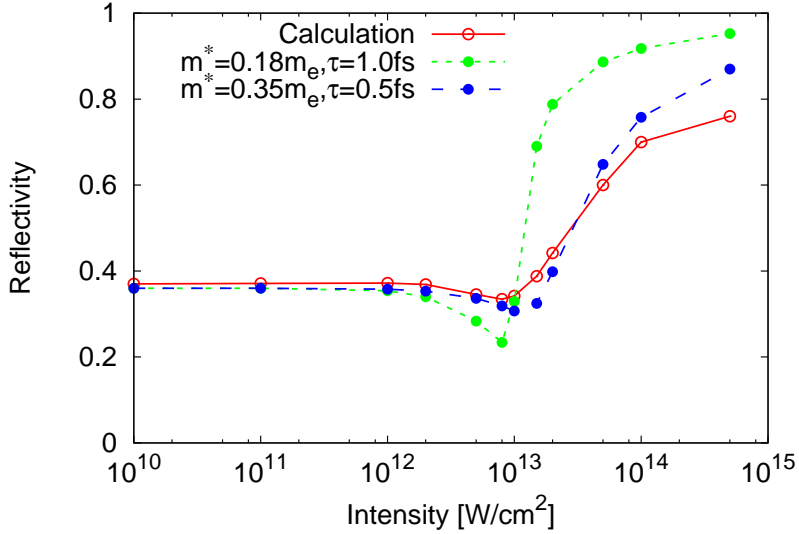}  
\caption{\label{r-fit}   
The reflectivity of Si at normal incidence is shown as a function of   
peak laser intensity. The red open circles with solid line repeat the 
calculated results from Fig. \ref{r}.  The green filled circles with dotted
line and blue filled circles with dashed line use Eqs. (\ref{R}) and 
(\ref{eps-model}) with $m^*=0.18m$, $\tau=1.0$fs,  and 
$m^*=0.35m$, $\tau=0.5$fs, respectively.
}  
\end{figure}
Figure \ref{r-fit} shows the comparison for two 
assumptions about the effective mass and Drude damping time.  
For a given laser intensity, we use the electron-hole density $n_{ph}$ 
in our calculation shown in Fig.~\ref{E-surface}. 
The red open circle with solid line is the present calculation.
The green filled circle with dotted line is the effective mass and
damping time adopted in Ref. \cite{so00}, $m^*=0.18m$ and
$\tau=1$ fs. The blue filled circle with dashed line is the
parameters adopted in Ref. \cite{me10}, $m^*=0.35m$ and
$\tau=0.5$ fs.
One can see that
on a qualitative level, both the dip and the strong increase can be
explained by plasma effects.  One could try to fit the plasma
parameters to reproduce the reflectivity curve, but it is probably not
realistic to assume that a fixed dielectric function is responsible for
the electromagnetic interactions.  However, it should be mentioned that
the reflectivity as well as the absolute value of the dielectric function
are minimized when the screened plasma frequency,
\be
\omega_p^2 = \frac{4 \pi e^2 n_{ph}}{\epsilon(\omega_{\ell}, 0) m^*},
\ee
coincides with the frequency of the incident laser pulse, 
$\omega_{\ell} = \omega_p$.
This relation is fulfilled at the laser intensity around
$10^{13}$ W/cm$^2$, consistent with the behavior of reflectivity.

In Figs.~\ref{I-dependence} and \ref{t-surface}, we observed an
emission of electromagnetic field following the main pulse of the
reflected wave at the laser intensity of $10^{13}$ W/cm$^2$. 
This phenomenon may also be understood with the model 
dielectric function. At this intensity, a small magnitude of
the dielectric function at the surface allows a penetration of 
transmitted wave inside the medium. However, the dielectric
function changes rapidly inside the medium due to the increase
of electron-hole pair density. It may cause a reflection from a 
deeper layer, producing the electromagnetic field following the
main pulse.

%
%

\subsection{Multi-scale vs single-cell approximations}

In Ref. \cite{ot08}, we calculated microscopic electron dynamics 
for an external electric field normal to the crystal surface and neglecting
magnetic fields.  In this longitudinal geometry,
the crystal response is uniform and the TDDFT is 
computationally much less expensive.  This was applied to the 
dielectric breakdown for diamond crystal, and the calculated 
threshold for 
breakdown was at least an order of magnitude higher than the measured 
threshold.

The aim of the present subsection is twofold. 
First we show that the present multi-scale calculation
gives much lower breakdown threshold than that of our previous 
calculation in the longitudinal geometry, thus resolving the discrepancy 
of our previous calculations with measurements.
Second, we clarify mutual relationship between the 
present multi-scale calculation and the single-cell treatment in either
the longitudinal or transverse geometry.
Since the single-cell calculations 
are much easier computationally, 
it would be useful to know what physical information 
may be extracted reliably from them.

\begin{figure}  
\includegraphics [width = 8cm]{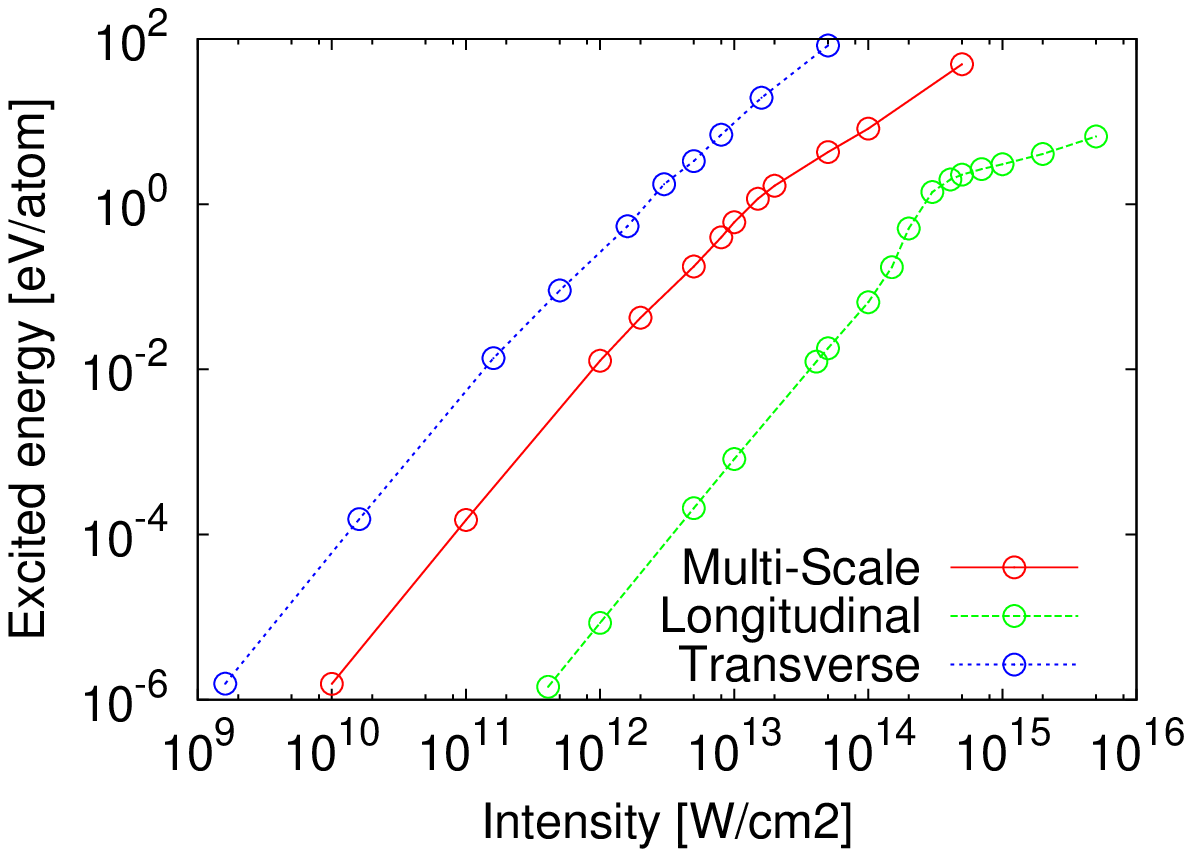}
\includegraphics [width = 8cm]{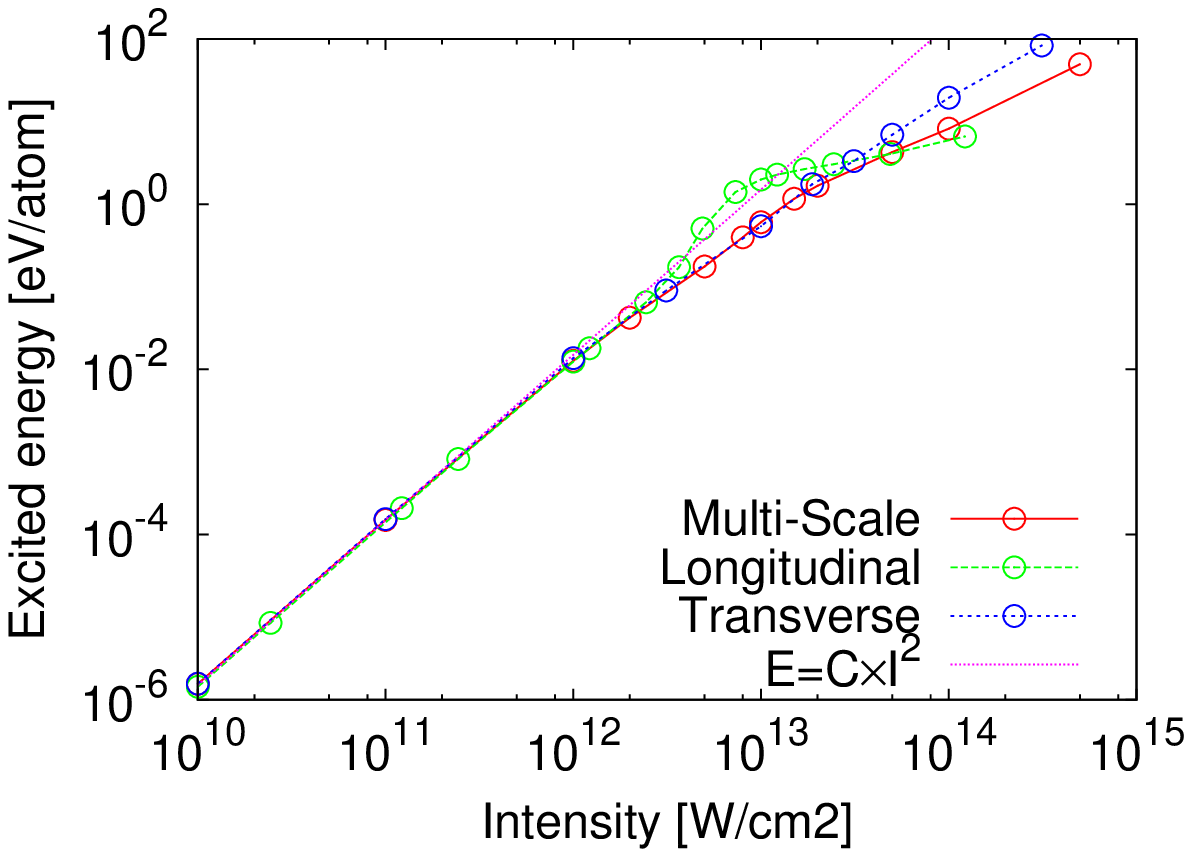}  
\caption{\label{e-comp} 
Deposited energy in the Si medium. Red solid line: the energy
deposited in the first-layer in the multi-scale calculation.
Green, dashed line: microscopic calculation in the longitudinal
geometry. Blue dotted line: microscopic calculation in the 
transverse geometry.
In the left panel, the horizontal axis is the intensity of the
incident laser pulse for multi-scale calculation, and is the
intensity of the applied laser pulses in the microscopic 
calculations of longitudinal and transverse geometries. 
In the right panel, the laser intensity is normalized to the 
transverse case. See the text for more detail.
}  
\end{figure}  
We first explain in more detail the longitudinal and
transverse geometries in the single-cell calculations.
In the transverse geometry, we simply put the vector
potential of applied laser pulse, $A(t)$, in the Kohn-Sham
Hamiltonian and calculate the electron dynamics.
In the longitudinal geometry, we take that field as external
and add to it the field from the induced current in the medium.
The vector potential
in the Kohn-Sham Hamiltonian is the sum of the external
and the induced fields, $A(t)=A_{ext}(t) + A_{ind}(t)$. 

The final-state electronic excitation energies for the three calculations are
shown in Fig. \ref{e-comp}.
In the left panel, the red circles and solid line shows the deposited
energy at the surface in the multi-scale calculation as a
function of the incident laser intensity.
The green circles and dashed line is the microscopic calculation
in the longitudinal geometry, as adopted in Ref. \cite{ot08}.
The blue circles and dotted line is the microscopic calculation
in the transverse geometry.
We may identify the dielectric breakdown at the laser intensity
where the electron excitation energy  per atom is about 1 eV.
One sees that the threshold for dielectric breakdown is very different
for the three calculations. The threshold is lower by an order of 
magnitude for the multi-scale and transverse cases than the 
longitudinal case.

The difference may be understood using a dielectric picture to relate
the internal and external fields.
In the transverse case, the applied electric field directly acts
upon electrons in the medium. In the case of the multi-scale 
calculation, the electric field in the medium and the incident field 
are related by
\be
{\cal E}_{medium} = {2\over 1 + \sqrt{\epsilon}}   {\cal E}_{in}
\ee
Putting the value of dielectric constant $\epsilon=16$,
the laser intensity is different between the transverse and
multi-scale calculations by a factor of $(2/5)^2=0.16$.
In the longitudinal case, in addition to the above factor
connecting medium and incident fields, we need to add
the following factor connecting the external and the medium fields,
\be
{\cal E}_{ext} = \epsilon {\cal E}_{medium},
\ee
The factor to correct the laser intensity is $16^2 (2/5)^2=41$
for the longitudinal geometry.
Taking these factors as corrections to the laser intensity, 
we replot the electronic excitation energy as a function
of laser intensity in the medium in the right panel of Fig.~\ref{e-comp}.
We see that these factors explain accurately the order-or-magnitude 
difference in the dielectric breakdown threshold.  
The electronic excitation energy coincides accurately below 
$10^{12}$ W/cm$^2$ for three calculations, where excitations 
are mostly by two-photon absorption.
There are some deviations around $10^{13}$ W/cm$^2$ and above,
where the resonant excitation is expected. The longitudinal
calculation shows an abrupt rise of the excitation energy which
we interpreted as a resonant energy transfer from the laser pulse
to the electrons \cite{ot08}. The other two calculations do not produce 
an abrupt rise but rather show a smooth saturation of the energy transfer. 

\section{Summary and outlook}

We have developed a first-principles framework to calculation the
propagation of electromagnetic field in crystalline solids. 
The macroscopic electromagnetic field is described by Maxwell
equations while the microscopic electron dynamics is described
by TDDFT.
With use of massively parallel computers, we showed that it is 
feasible to treat one of the simplest systems of physical interest,
the propagation of a laser pulse into bulk Si 
at the normal incidence.

At low field intensity, the calculated field propagation and electronic
excitations exhibit features expected from ordinary electromagnetic
theory
with the dielectric function given by linear response theory.
The electronic excitations are dominated by two-photon absorption 
at low intensities since the laser frequency is below the direct 
bandgap.

As the laser intensity increased, the density of excited electron-hole 
pairs become high enough to affect the response.  This is conveniently
modeled as an electron-hole plasma. 
At around $10^{13}$ W/cm$^2$, the plasma frequency of
excited electrons reaches the visible frequency, showing a 
nonlinear interaction with the incident laser pulse. Above this 
intensity, the responses are dominated by nonlinear electron dynamics.

We have also found that the surface absorption obtained in the 
multi-scale theory can be 
described by a single-cell approximation using dielectric formulas
to relate the internal and external fields, provided the fields do 
not much exceed $10^{13}$ W/cm$^2$.

Finally, we mention some directions that might be interesting to 
take up in later work.  Analytic approximations have been proposed 
to express the excitation energy as a function of the Keldysh parameter
\cite{ke65}.  We have not examined the validity or accuracy of such
approximations, but it would be useful to have this information.

Computations in the present framework could be extended to deal with
laser pulses at oblique angles of incidence.  In that case, the 
field is not translationally invariant in the $x$ direction, but the
medium itself is.  Consequently relatively few cells would be needed to describe the
$x$ dependence.  It would also be interesting to extend the present
calculations to pump-probe laser pulse protocols.  In principle it
is straightforward to calculate the response to a double pulse separated
in time.  As a practical matter, pump-probe responses could most easily
be studied in a single-cell approximation.  Also, one could examine
the linear response of the excited system using fields of the
pump-probe form.  This is important to verify the validity of the
arguments made in Sect. \ref{surface-layer}.

\section*{Acknowledgment}  
 
This work is supported by the Grant-in-Aid for Scientific Research
Nos. 23340113, 23104503, 21340073, and 21740303. The numerical  
calculations were performed on the supercomputer at the Institute  
of Solid State Physics, University of Tokyo, and T2K-Tsukuba at
the Center for Computational Sciences, University of Tsukuba.
GFB acknowledges support by the  
National Science Foundation under Grant PHY-0835543 and by the   
DOE grant under grant DE-FG02-00ER41132.  
  
\section*{Appendix}
Here we show how the dielectric function may be calculated using
the formalism of Sect. II.E.
For the perturbation, we take $A$ to be of the form
\be
\label{A_for_eps}
A(t) = A_0 \theta(t)\,.
\ee
The microscopic equation of motion Eq.~(\ref{TDKS}) is integrated 
from $t=0$ to $t=T_m$ to obtain the $J(t)$ over the time interval.  
As is evident from Eq.~(\ref{A2J}), the calculated $J(t)$ is
proportional to the conductivity as a function of time,
\be
\sigma(t) = -\frac{c}{A_0}J(t).
\ee
This is Fourier transformed as
\be
\sigma(\omega) = \int_0^{T_m} dt \,e^{i\omega t} f(t)\sigma(t)
\ee 
where $f(t)$ is a filter to suppress spurious oscillations that would
arise from a sharp cutoff of the integration at $T_m$.
We employ a third order polynomial for it \cite{ya06}.
The dielectric function may be obtained from the conductivity by
Eq.~(\ref{eps-sgm}).

\begin{figure}  
\includegraphics [width = 11cm]{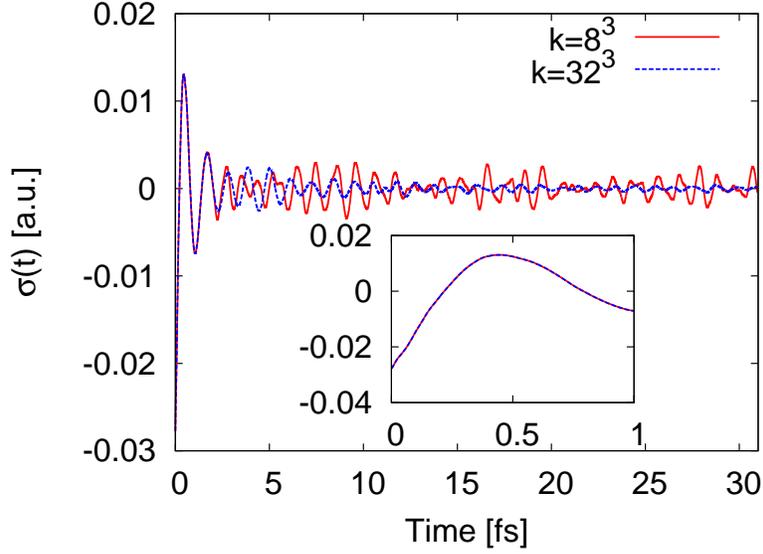}  
\caption{\label{sigma-t} 
Conductivity as a function of time.
Calculations for two choices of $k$-points are compared.
}  
\end{figure}  
We carried out this computation taking $A_0=0.0005$ a.u. and 
$T_m=31$ fs (16,000 time steps with $\Delta t=0.08$ a.u.).  
In Fig.~\ref{sigma-t}, we show a conductivity as a function of time,
$\sigma(t)$, for two choices of $k$-points, $8^3$ and $32^3$.
In our multi-scale calculation, we adopt $8^3$ $k$-points.
Two calculations coincide each other up to 2 fs. 
There remain oscillations for a long period in the calculation 
of $8^3$ $k$-points, which are washed out if one employs
a finer $k$-points grid.

\begin{figure}  
\includegraphics [width = 11cm]{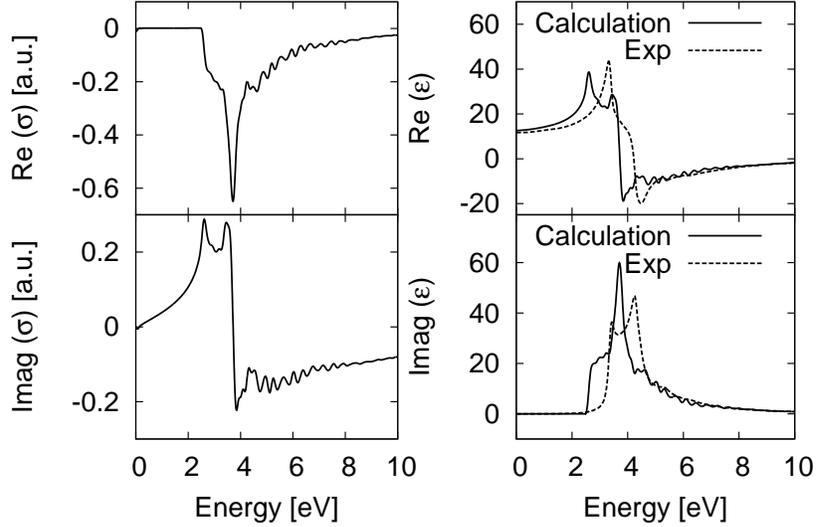}  
\caption{\label{sigma-eps} 
Conductivity and dielectric function as a function of frequency
in which $32^3$ $k$-points are used.
The measured value is also shown for dielectric function.
}  
\end{figure}  
The conductivity $\sigma(t)$ is Fourier transformed to obtain
the conductivity and dielectric function as a function of
frequency. They are shown in Fig.~\ref{sigma-eps}, in which
$32^3$ $k$-points are used.
At a frequency region close to zero, the conductivity $\sigma(\omega)$
should behave
\be
\sigma(\omega) = i \left. \frac{d\sigma}{d\omega} \right \vert_{\omega=0}
\omega.
\ee
In actual calculation, it is not exact due to the presence of spurious mode
which originates from a violation of translational invariance in the
real-space grid calculation.
Since a small deviation from the above analytic behavior at around
$\omega=0$ harms the low frequency behavior of the dielectric
function, we replace the real part of the dielectric function by a
second order polynomial of the frequency below 1 eV.
The calculated dielectric function, $\epsilon(\omega)$, shown in 
the right panels of Fig.~\ref{sigma-eps}, is very close to the one 
calculated in Ref. \cite{sh10} using the formalism of Ref. \cite{be00}.

The calculated real part of the static dielectric function is 
$\epsilon(0)=12.6$, close to the experimental value of 11.6.
However, as is well known in the density functional theory, 
the direct band gap in the local density approximation is smaller than the
experimental one (2.4 eV theory vs. 3.1 eV experiment).

\end{document}